\documentclass{llncs}
\usepackage[utf8]{inputenc}
\usepackage{graphicx}
\usepackage[english]{babel}
\usepackage[T1]{fontenc}
\usepackage{amsmath}
\usepackage{amssymb}
\usepackage{bbm}
\usepackage{subfig}
\usepackage{color}
\usepackage{slashbox}
\usepackage{algorithm,algorithmic}

\usepackage{todonotes}
\usepackage{marginnote}
\usepackage{hyperref}
\hypersetup{colorlinks=false,
			pagebackref=true,draft=false}

\newcommand{\ml}{mailing-list}
\newcommand{\debianml}{Debian \ml}

\usepackage{mathptmx}       
\usepackage{helvet}         
\usepackage{courier}        
\usepackage{type1cm}        
\usepackage{makeidx}         
\usepackage{graphicx}        
\usepackage{multicol}        
\usepackage[bottom]{footmisc}
\usepackage{url}

\begin{document}

\mainmatter  

\title{Analysis of the temporal and structural features of threads in a mailing-list}
\titlerunning{Analysis of the temporal and structural features of threads in a mailing-list}

\author{No\'{e} Gaumont\inst{2}, Tiphaine Viard\inst{2}, Raphaël Fournier-S'niehotta\inst{1}, Qinna Wang, and Matthieu Latapy \inst{2}}

\authorrunning{N. Gaumont, T. Viard, R. Fournier-S'niehotta, Q. Wang and M. Latapy} 

\institute{
CNAM, CEDRIC
\email{fournier@cnam.fr},\\
\and
Sorbonne Universités, UPMC Univ Paris 06, CNRS, LIP6 UMR 7606, 4 place Jussieu 75005 Paris
\email{first.last@lip6.fr}
}
\toctitle{Analysis of the temporal and structural features of threads in a mailing-list}
\tocauthor{N. Gaumont et al.}
\maketitle

\begin{abstract}{A link stream is a collection of triplets $(t,u,v)$ indicating that an interaction occurred between $u$ and $v$ at time $t$.
Link streams model many real-world situations like email exchanges between individuals, connections between devices, and others. 
Much work is currently devoted to the generalization of classical graph and network concepts to link streams.
In this paper, we generalize the existing notions of intra-community density and inter-community density.
We focus on emails exchanges in the Debian \ml\, and show that threads of emails, like communities in graphs, are dense subsets loosely connected from a link stream perspective.}
\end{abstract}

\section{Introduction}

Exchanges in a \ml\ are often studied as complex networks: there is a link between two individuals if they exchange emails. In particular, communities in such complex networks capture groups of friends or close colleagues (individuals that exchange many more messages within the group than outside the group, typically)~\cite{fortunato2010}. However, removing all time information has important consequences if one wants to study the dynamics of email exchanges.

In order to study those dynamics, one may label each link with the frequency of exchanges or the times at which they occur \cite{sun2007}, but capturing both the structure and the dynamics of exchanges remains challenging. In particular, studying threads calls for methods that capture the temporal nature of interactions more accurately, without loosing the power of network analysis.

\begin{figure}[!h]
\centering
\includegraphics[width=0.9\linewidth]{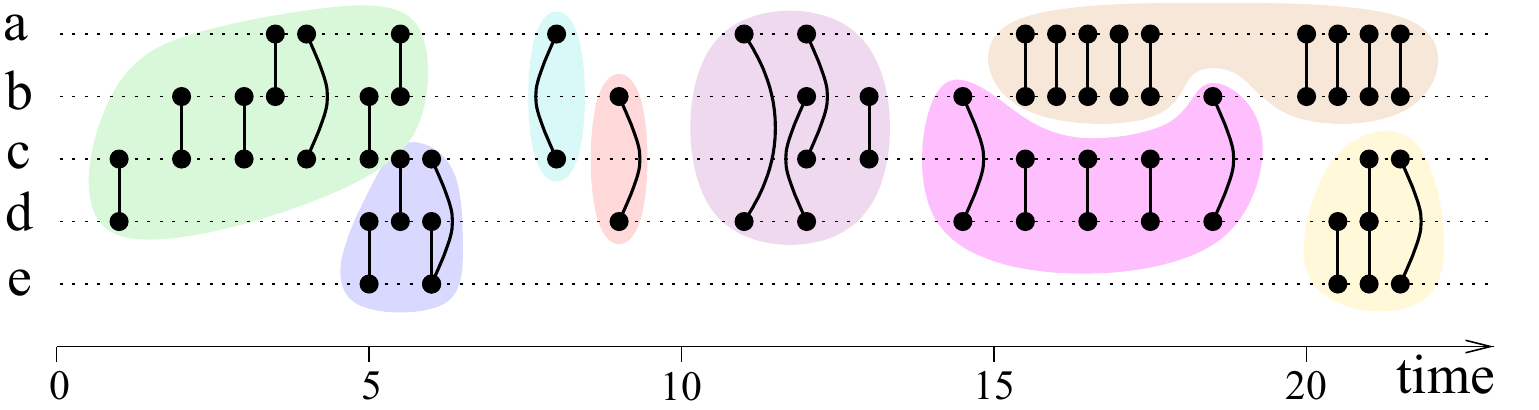}
\caption{An example of link stream representing email exchanges between individuals $a$, $b$, $c$, $d$ and $e$, with threads represented by colored areas. For instance, at time $5$, $b$ and $c$ exchange an email, as well as $d$ and $e$. Threads are {\em a priori} dense series of exchanges involving a limited group of nodes during a limited period of time.
}
\label{fig:threads-in-ls}
\end{figure}

We propose here to model email exchanges directly as link streams, {\em i.e.} series of triplets $(t,a,b)$ meaning that individuals $a$ and $b$ exchanged an email at time $t$. We then introduce notions that capture both the temporal and structural nature of these exchanges. We use a typical dataset obtained from a public mailing-list archive to illustrate our approach. We analyze this dataset using our model, with a special focus on the properties of threads within the whole archive. Our goal is to understand how the now classical concept of communities in complex networks may translate to threads in link streams representing email exchanges. Indeed, we expect the exchanges of a given thread to involve a specific set of individuals for a specific period of time, thus being dense from both structural and temporal points of view. This is illustrated in Figure~\ref{fig:threads-in-ls}.

\section{Dataset}

Archives of exchanges in various mailing-lists are readily available on the web, and studying them provides very rich insights on various issues. They have the advantage of being publicly available in many cases, and some involve large amounts of users over long time periods.

A typical example is provided by \debianml~\cite{debianml}: it contains emails sent from over $51753$ email addresses, over almost $20$ years. In addition, exchanges in this \ml\ have been studied in the past \cite{dorat2007,sowe2006,wang2014}. Finally, this dataset provides the thread information for each message, that we can use as a ground truth. For all these reasons, we use in this paper the \debianml\ to illustrate and validate our approach.

More precisely, we crawled the \debianml\ web archive \cite{debianml}.
For each message $m$, we extract its author $a(m)$, the date $t(m)$ at which it was posted (converted into \textsc{UTC} time), and the message it is replying to $p(m)$ (through the \textsc{in-reply-to} entry), which has a corresponding author $a(p(m))$. This corresponds to an interaction between $a(m)$ and $a(p(m))$ at time $t(m)$ in the link stream.
Some messages are not answers to any other message (they are directly sent to the \ml), and in this case we state that $p(m)=m$. Such messages are called {\em root} messages.

We capture the mailing-list from January 1st, 1996 to December 31st, 2014.
We obtain a dataset $\mathcal{D}$ of $n=722716$ emails sent from $51753$ distinct email addresses. 

Each root message $m$ naturally induces a thread: it is the set $\mathcal{T}(m)$ of messages such that $m$ belongs to $\mathcal{T}(m)$ and if a message $m'$ is in $\mathcal{T}(m)$ then all messages $m''$ such that $p(m'')=m'$ also belong to $\mathcal{T}(m)$. In other words, $\mathcal{T}(m)$ contains exactly $m$, the answers to $m$, the answers to these answers, and so on. The focus of this paper is the study of structural and temporal features of these threads.

Our data contains incomplete threads: the ones that have an email in our dataset but began before and/or continued after the data collection period. Some threads also exhibit inconsistencies, for instance a reply has a smaller timestamp than the message it replies to. We remove those threads, as well as all threads that last for more than $2$ years, or that start $2$ years before the end of our data collection.


After this bias correction procedure, we obtain $n=554233$ emails, involving $34648$ distinct authors over a duration of $598532269$ seconds ($18$ years, $11$ months and $19$ days) and $116999$ threads. 

\section{Framework and notations}

Our goal is to study the structural and temporal properties of threads within a mailing-list archive. In order to do so, we propose a model of the data that captures both its temporal and structural nature, and allows for easy manipulation of threads.


We model our mailing-list archive as the link stream $D = (T_D, V_D, E_D)$ with $T_D = [\alpha,\omega]$, $V_D = \{a(m): m\in \mathcal{D'}\}$ and $E_D = \{(t(m), a(m),a(p(m))): m\in \mathcal{D'}\}$ where $\mathcal{D'}$ is the set of emails in our dataset after cleaning. In other words, a triplet $(t, u, v)$ in $E_D$ indicates that individual $u$ answered to an email of individual $v$ at time $t$.


Such a link stream naturally contains sub-streams: $L' = (T', V', E')$ is a substream of $L = (T,V,E)$ if and only if $T'\subseteq T$, $V'\subseteq V$ and $E'\subseteq E$. In other words, all the interactions of $L'$ also appear in $L$. Given a set of nodes $S$, we define the sub-stream $L(S)$ of $L$ induced by $S$ as the largest sub-stream of $L$ such that all the links in $L(S)$ are between nodes in $S$.


Any link stream $L=(T,V,E)$ also induces a graph $G = (V_G, E_G)$ where $V_G = \{u: \exists t\in T, v\in V$ s.t. $(t,u,v) \in E\}$ and $E_G = \{(u,v): \exists t\in T $ s.t. $(t,u,v)\in E \}$. In our case, the whole mailing-list archive induces the graph $G(D)$ among authors of emails, and each thread induces a sub-graph of $G(D)$.

In a graph $G=(V,E)$, a community structure is defined by a partition $C = \{C_i\}_{i=1..k}$ of $V$ into $k$ communities. In other words, $\bigcup_i C_i = V$ and $C_i \cap C_j = \emptyset$ whenever $i\not=j$.
In a similar way, one may consider a link stream $L = (T,V,E)$ and a partition of its links into $k$ sub-streams $P = \left\{P_i = (T_i, V_i, E_i)\right\}_{i=1..k}$. In other words, for any $(t,u,v)\in E$, there exists a unique $j$ between $1$ and $k$ such that $(t,u,v)$ is a link of $E_j$.

The threads in our email dataset are exactly a partition of the whole stream, which we denote by $\mathcal{T} = \{P_i\}_{i=1..k}$ where $k$ is the number of threads and each $P_i$ is a sub-stream representing a thread (with our notations above, there exists a message $m$ such that $P_i = \mathcal{T}(m)$). See Figure~\ref{fig:threads-in-ls}.

Notice that, although the threads are a partition of the whole stream, their induced graphs may overlap: some nodes and links of $G(D)$ belong to several sub-graphs $G(P_i)$. As a consequence, threads do not induce a partition of $G(D)$ into communities. Instead, one may see the partition of $D$ into threads as a community structure, and this is the focus of our work.



Notice finally that we consider that links are undirected (i.e. $(t,u,v) = (t,v,u)$) and happen at an instant in time (regardless, for instance, of when the message is read).
Taking into account the direction and duration of links is out of the scope of this work.

\section{Basic statistics}

In this section, we present the basic statistics describing the threads in our dataset and the whole archive.

\begin{figure}
	\includegraphics[angle=-90, width=0.49\linewidth]{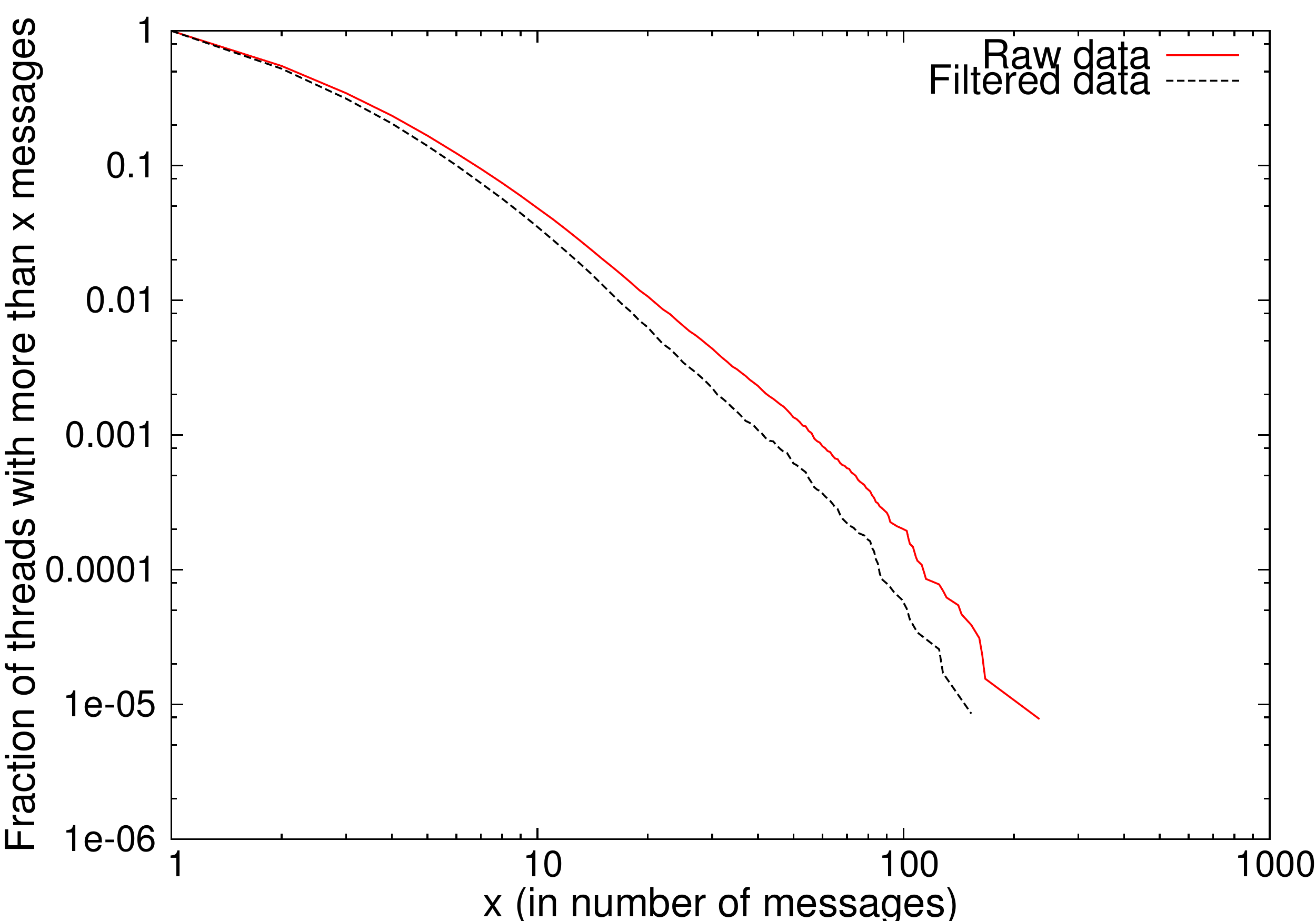}
	\includegraphics[angle=-90, width=0.49\linewidth]{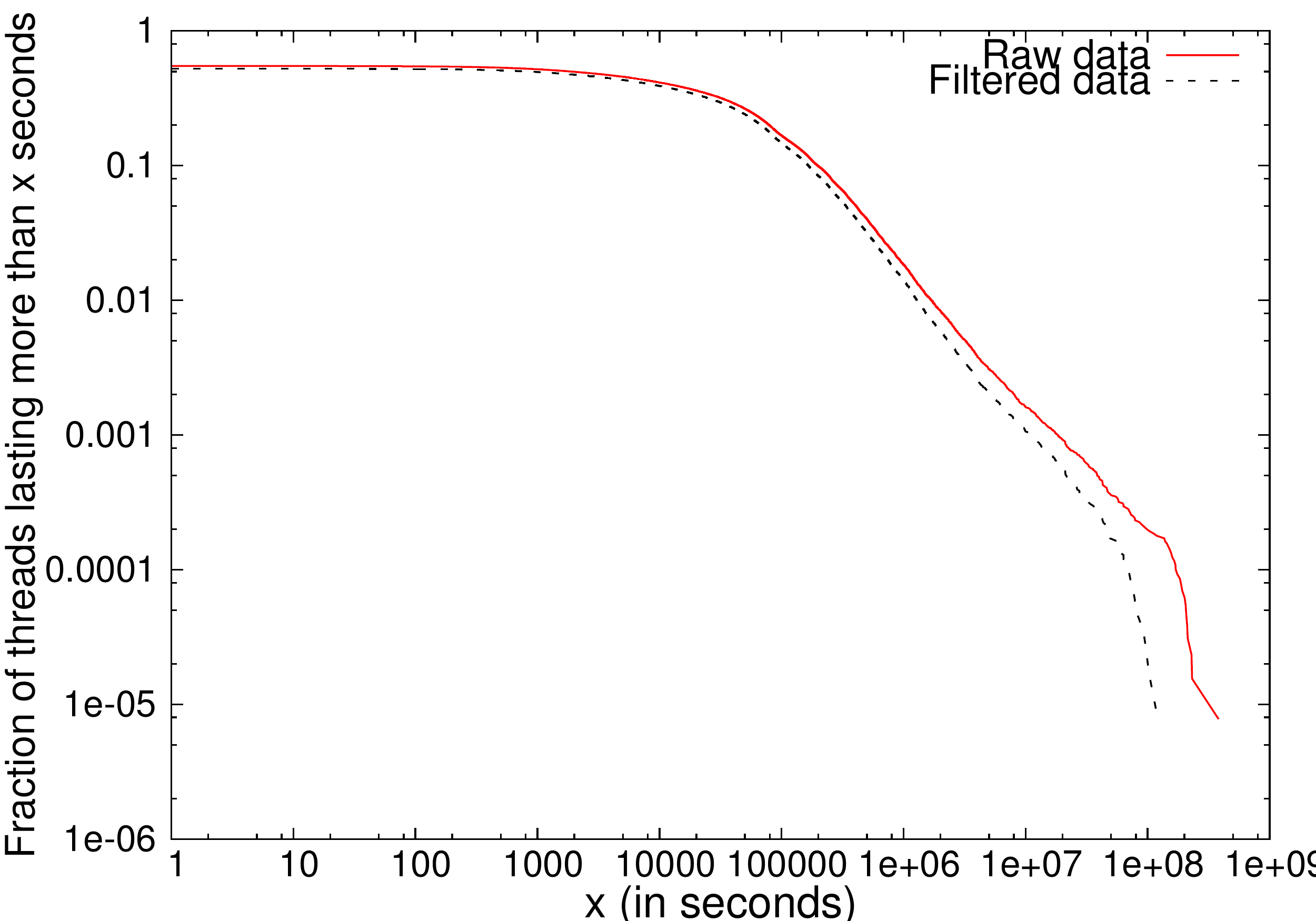}\\
	\includegraphics[angle=-90, width=0.49\linewidth]{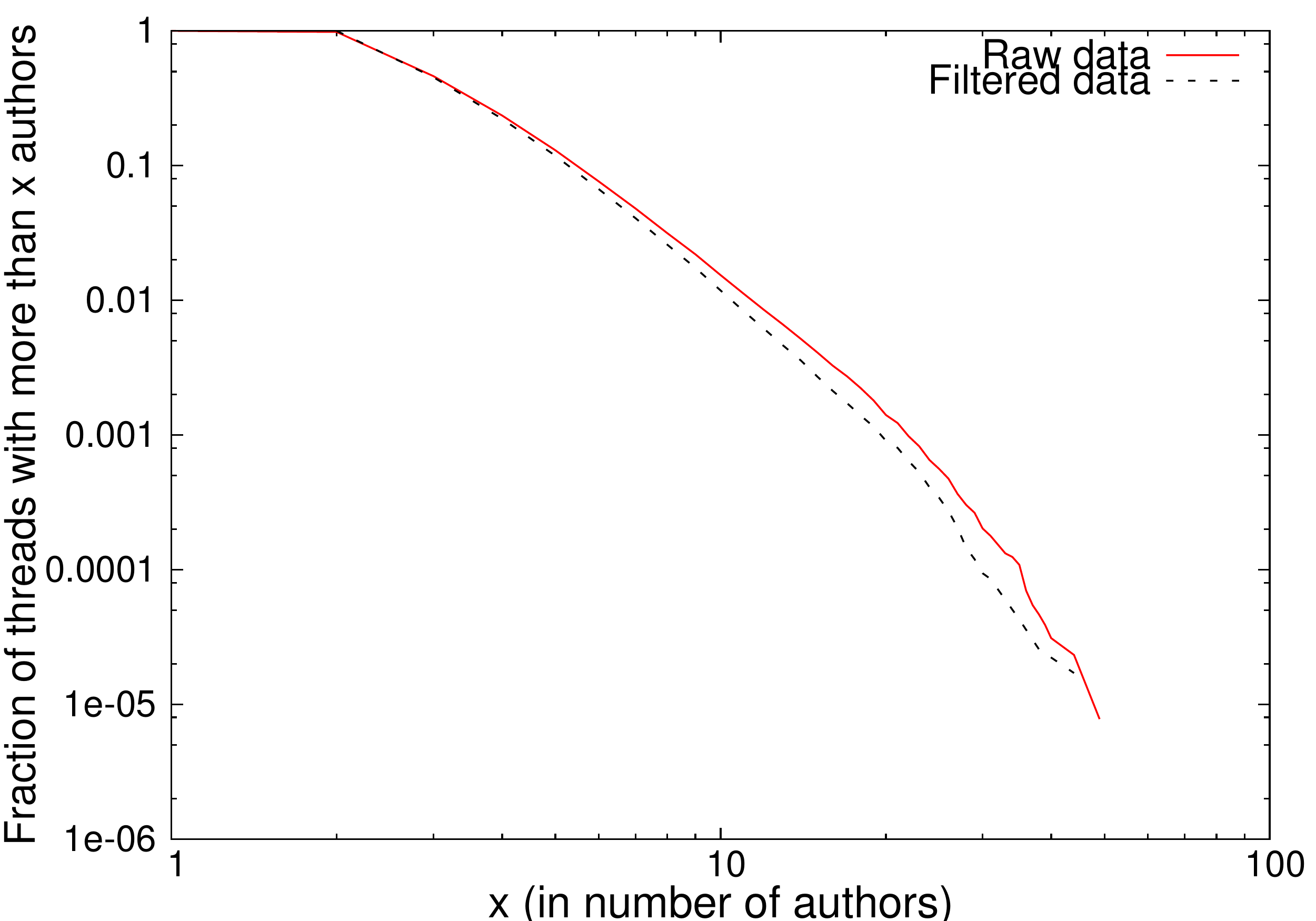}
	\includegraphics[angle=-90, width=0.49\linewidth]{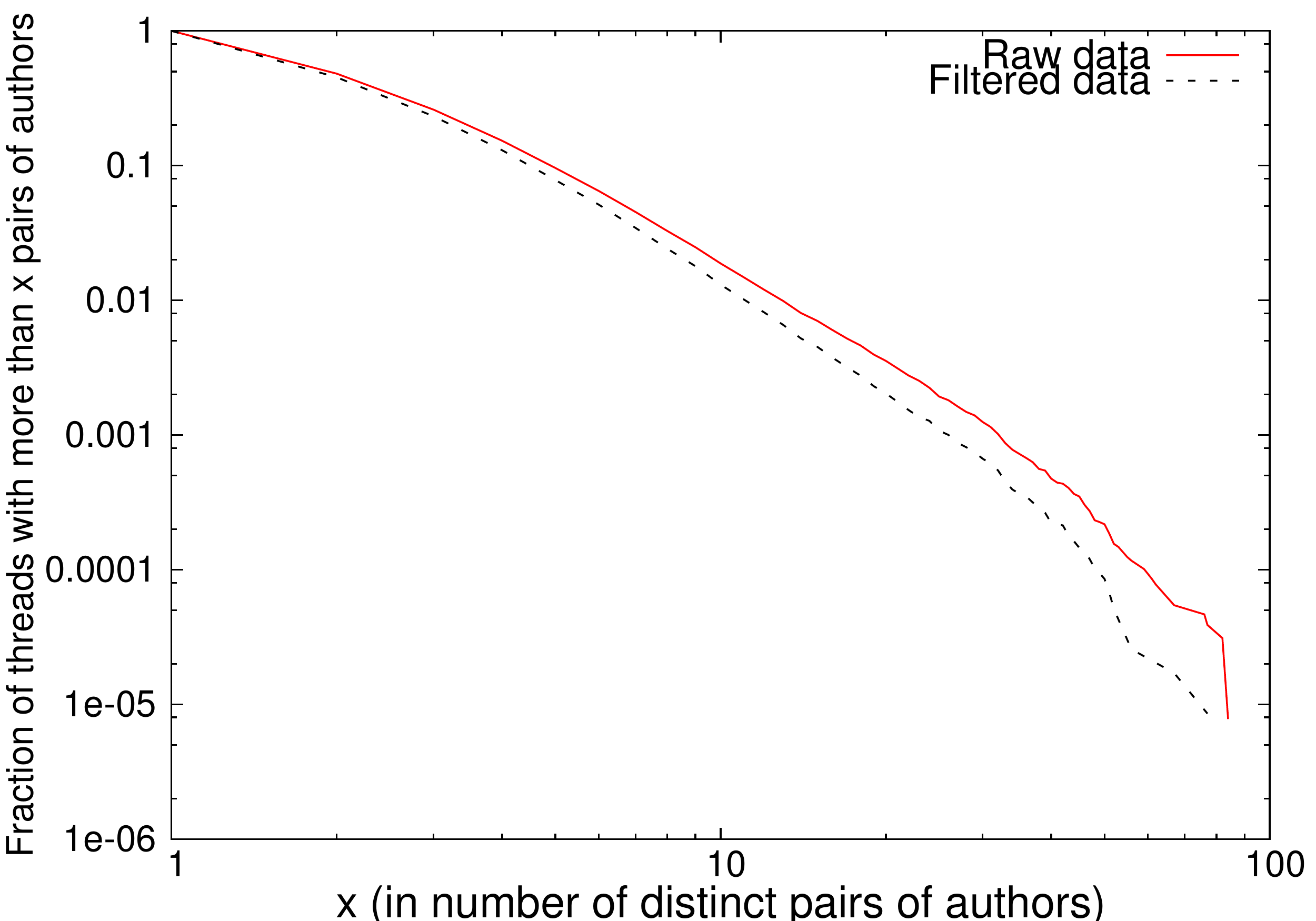}
	
	\caption{Complementary cumulative distributions for basic statistics of our raw (solid line) and filtered (dotted line) datasets. Top left: thread sizes (number of messages per thread); top right: thread durations (time elapsed between the first and the last message of the thread); bottom left: number of distinct authors; bottom right: number of distinct pairs of authors.}
	\label{fig:dists}
\end{figure}

The most basic description of our data certainly is the number of links ({\em i.e.} emails) they contain, the number of distinct nodes ({\em i.e.} authors) involved, the number of distinct links they contain (distinct pairs of authors in direct interaction), and their duration (time from the first email to the last one).
Figure~\ref{fig:dists} display the distribution of these values for each thread.

Although the largest thread lasts more than a year, most threads are contained within a few days ($100000$ seconds is a bit more than $24$ hours). Similarly, the largest thread involves $100$ messages, though all intermediate sizes are represented in the dataset. Most threads are very short and involve less than $3$ messages. 

\begin{figure}
	\includegraphics[angle=-90, width=0.49\linewidth]{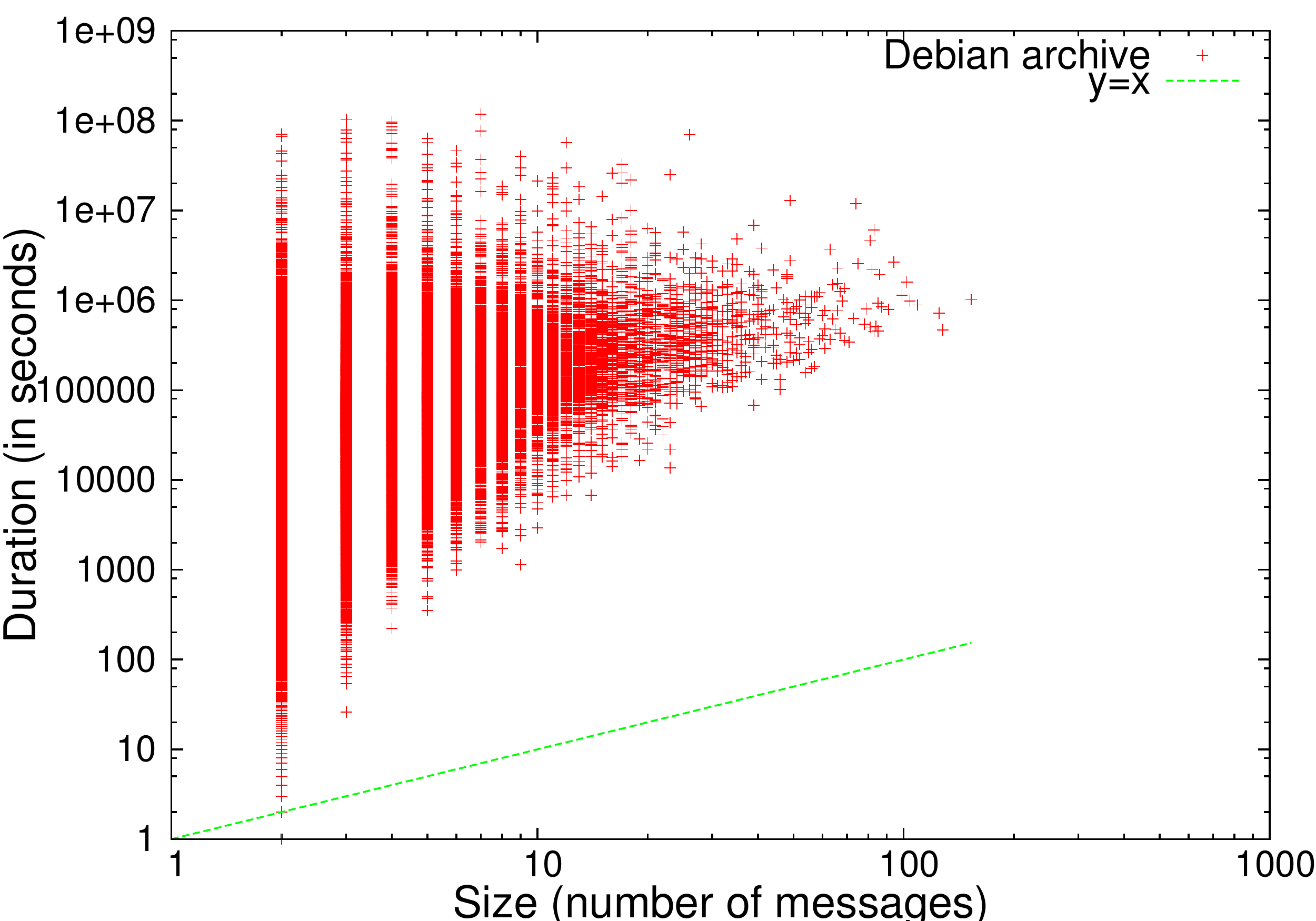}
	\includegraphics[angle=-90, width=0.49\linewidth]{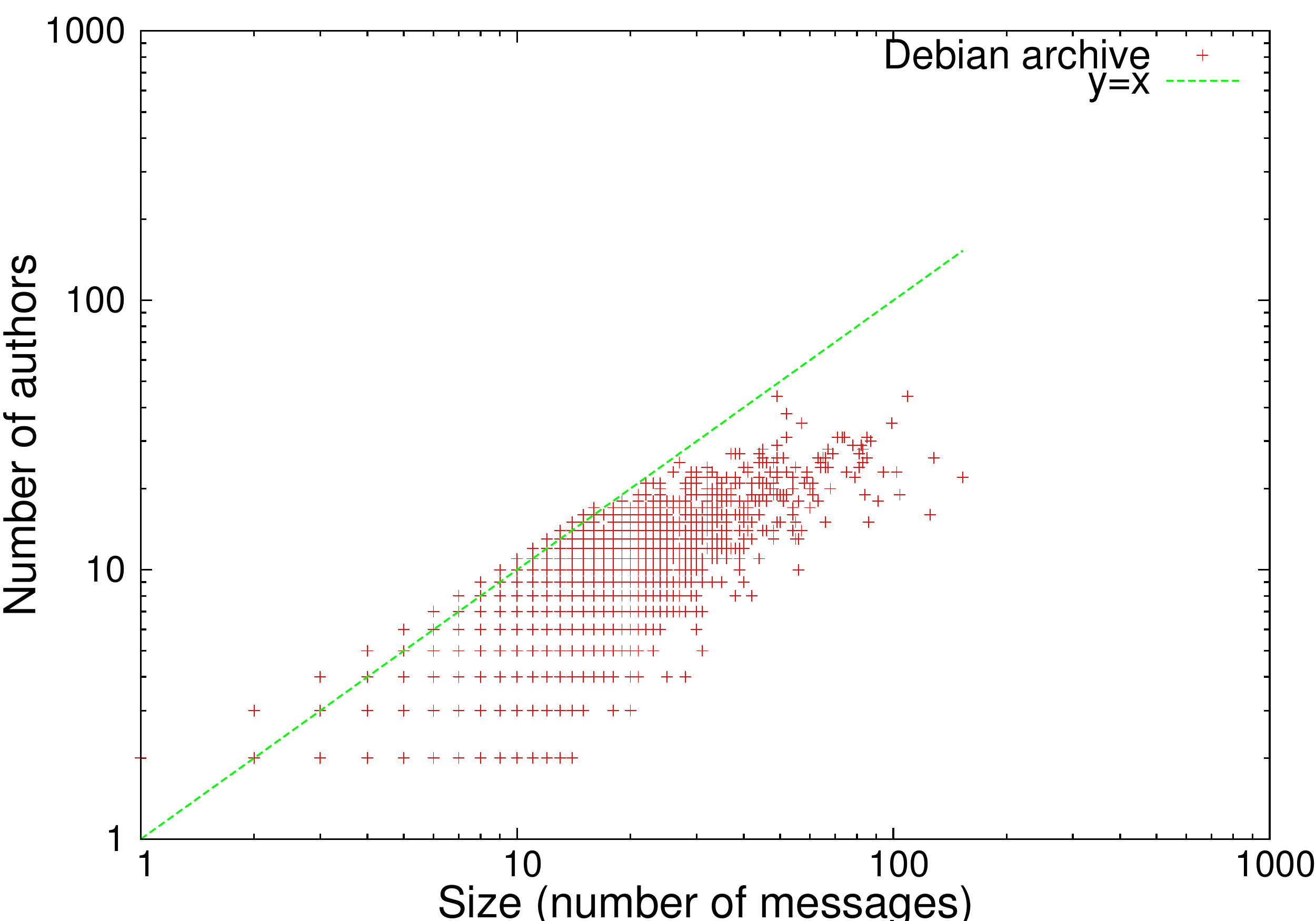}
	\caption{Left: Correlations between size and duration of threads. Right: Correlations between size of threads and the number of authors involved.}
	\label{fig:corr}
\end{figure}

In order to gain more insight, we observe correlations between some of these basic statistics. Figure~\ref{fig:corr} (left) shows that thread duration and size are correlated (the larger a thread is, the longer it  is likely to be); notice however that for small-sized threads, all types of durations are represented. Looking at the correlations between the size of threads and the number of distinct authors involved  shows that threads nearly always involve more messages than authors. This is a typical feature of \ml{}s~\cite{dorat2007} and as such is dataset-dependent.

In a link stream $L = (T,V,E)$ with $T = [\alpha,\omega]$, we define, for all $(u,v) \in V\times V$, the maximal sequence $t_{uv} = (\alpha, t_0, \dots, t_k, \omega)$ such that for all $i$ between $0$ and $k$, there exists $(t_i, u, v)\in E$, and for all $i$ between $0$ and $k-1$, $t_i \leq t_{i+1}$. In other words, $t_{uv}$ is the ordered sequence of apparitions of the link $(u,v)$ to which we add $\alpha$ and $\omega$.

\begin{figure}
	\centering
	\includegraphics[width=0.49\linewidth]{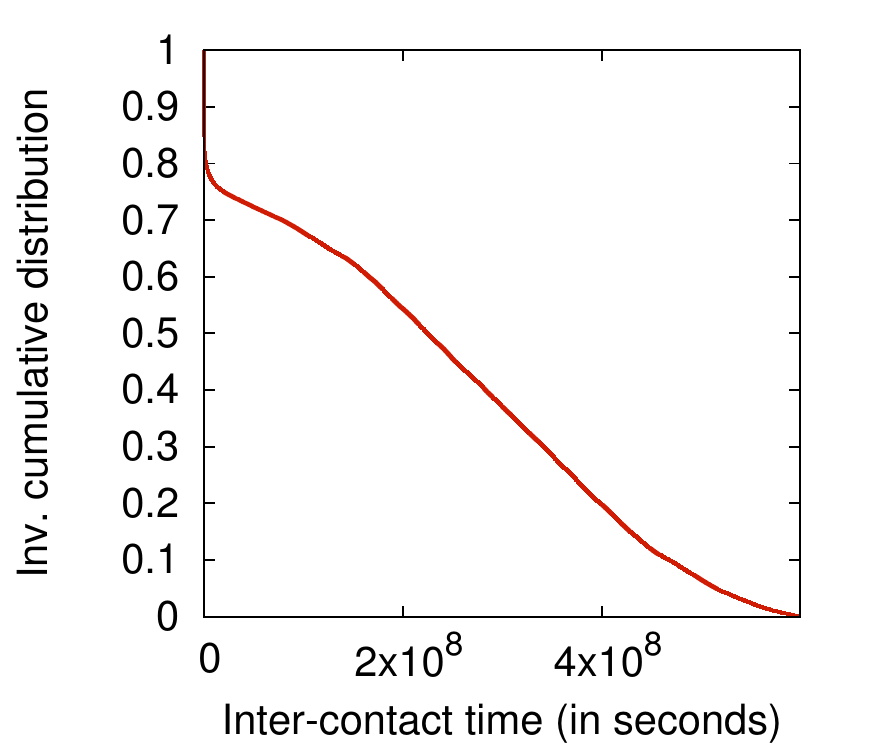}
	\includegraphics[width=0.48\linewidth]{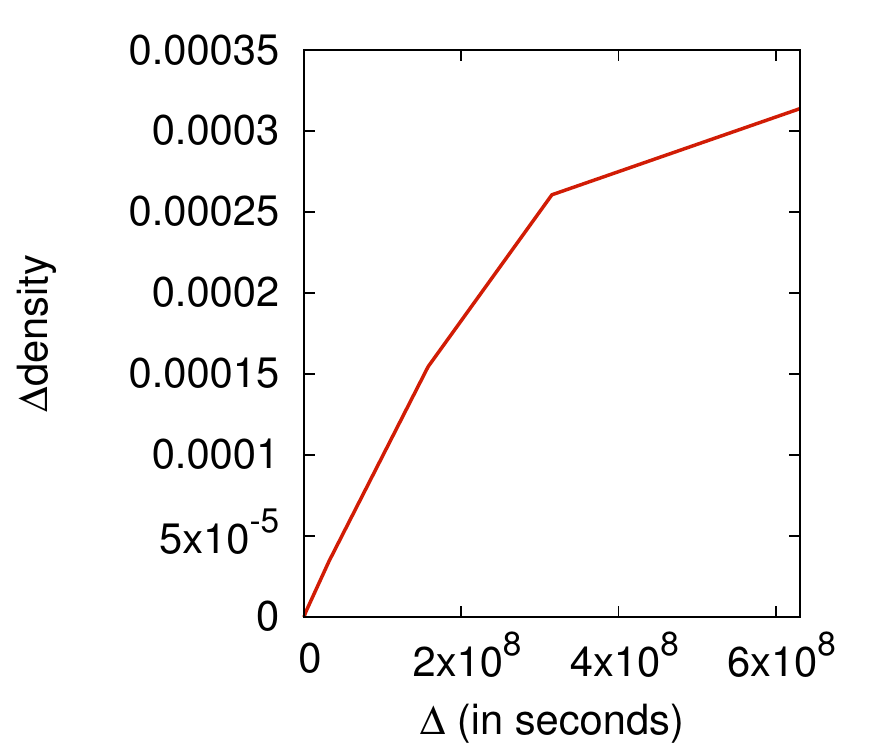}
	\caption{Left: Inter-contact times distribution in the \debianml{} dataset. Right: Evolution of the $\Delta$-density of the link stream for $\Delta$ from $ 1$ second to $20$ years.}
	\label{fig:ict}
\end{figure}

We further define $\tau(u,v) = (t_{i+1}-t_i)_{i=0..k+1}$ the sequence of intercontact times of a pair of $u$ and $v$ in $V$. 
In other words it is the series of times elapsed between two consecutive occurrences of a link between them.
Figure~\ref{fig:ict} (left) shows the inter-contact times distribution in the \debianml{} for all pairs of nodes $(u,v)$. 

\section{Interactions within threads}

The key feature of communities is the fact that they form dense subgroups. This section is therefore devoted to the study of density of interactions within threads, from both structural and temporal point of views.

\subsection{Density of threads}

In a graph, the density is the probability that two randomly chosen nodes are linked together. In other words, it captures the extent at which {\em all} nodes are directly connected to each other. The density of the graph $G(D)$ induced by our dataset is $3.139\times 10^{-4}$. 

In \cite{viard2014}, we introduced the notion of $\Delta$-density to capture a similar intuition in link streams, involving both structure and time. Indeed, given a duration $\Delta$, the $\Delta$-density of link stream $L$ is the probability that a link appears between two randomly chosen nodes during a randomly chosen time interval of duration $\Delta$. It captures the extent at which all nodes are directly connected to each other at least every $\Delta$ time units. Formally, it is defined as:
$$
\delta_\Delta(L) = 1 - \frac{2\cdot \sum_{u,v\in V, u\not= v} \sum_{t\in\tau(u,v)} \max(0,t-\Delta)}{|V|\cdot (|V|-1)\cdot\max(0,\omega-\alpha-\Delta)}
$$
where $\tau(u,v)$ denotes the inter-contact times between $u$ and $v$, and $\alpha$ and $\omega$ are the start and end time of the link stream.

In order to study the $\Delta$-density in our data, we first have to choose an appropriate $\Delta$. We use here several values which capture email dynamics at different scales: $\Delta = $ 1 minute, 1 hour, 1 day, 1 week, 1 month, 1 year and 20 years (the whole duration of the dataset). Figure~\ref{fig:ict} (right) displays the evolution of the $\Delta$-density of the stream for all theses values of $\Delta$. It shows that the $\Delta$-density is small for small $\Delta$s, and converges to the density of the graph induced by the email exchanges (in our case, $3.139\cdot 10^{-4}$). 

In Figure~\ref{fig:ict} (right), the inflexion points give information on the values of $\Delta$ where the dynamics change. Still, looking at the density of the whole stream is very coarse and yields little information. A finer approach consists in looking at the $\Delta$-density of relevant sub-streams. In our case, the threads between authors are a natural object to study. 

\subsection{Intra-thread density}

More globally, given a graph $G=(V,E)$ and a partition $C = \{C_i\}_{i_1..k}$ of $V$ into $k$ communities, the density within communities of $C$ is captured by the {\em intra-community density}:
$$
\frac{2\cdot\sum_i|\{(u,v) \in E\mbox{, }u \in C_i \mbox{ and } v \in C_i\}|}{\sum_i|C_i|\cdot(|C_i|-1)}
$$
In other words, intra-community density is the probability that two nodes chosen at random in the same community are linked together.

In our case, this notion does not directly make sense: as already noticed, we do not have communities defined on $G(D)$ since the graphs induced by threads overlap. However, we extend the notion of intra-community density to link streams as follows. The intra-thread $\Delta$-density is the probability that two randomly chosen authors contributing to the same thread are linked together within a randomly time interval of duration $\Delta$, for a given $\Delta$:
$$
1 - \frac{2\cdot \sum_i \sum_{u,v\in V_i, u\not= v} \sum_{t\in\tau_i(u,v)} \max(0,t-\Delta)}{\sum_i |V_i|\cdot (|V_i|-1)\cdot\max(0,\omega_i-\alpha_i-\Delta)}
$$
where $V_i$ is the set of authors involved in thread $P_i$, $\alpha_i$ is the time of the first message in the thread (i.e the minimal $t$ such that there exists a $(t,u,v)\in E_i$), $\omega_i$ is the time of the last message in the thread (i.e the maximal $t$ such that there exists a $(t,u,v)\in E_i$) and $\tau_i(u,v)$ denotes the inter-contact times in $P_i$.

\begin{figure}
\centering
	\includegraphics[width=0.48\linewidth]{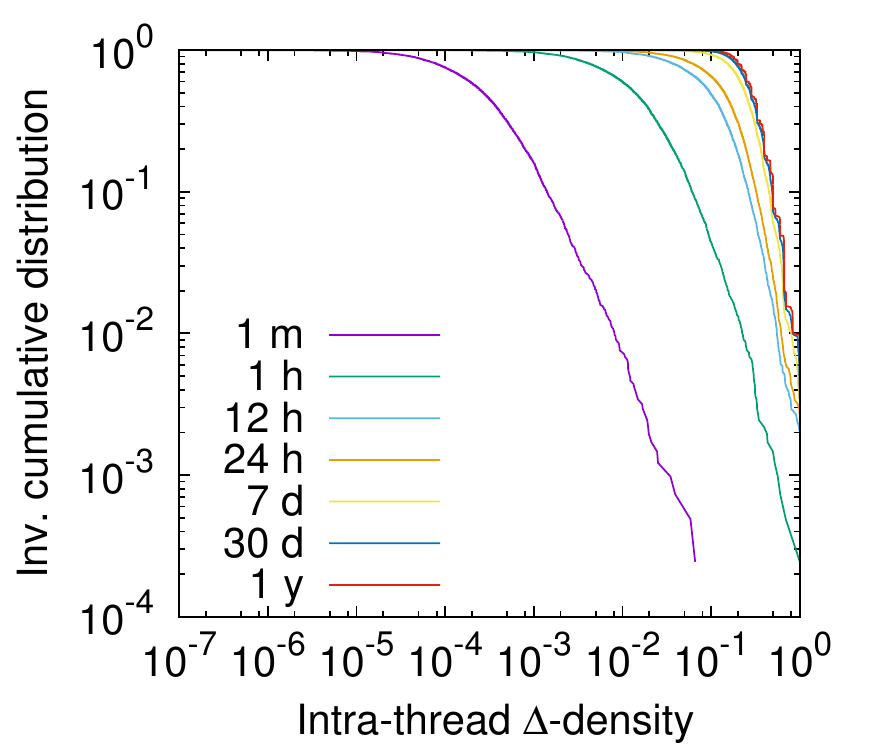}
\hfill
	\includegraphics[width=0.48\linewidth]{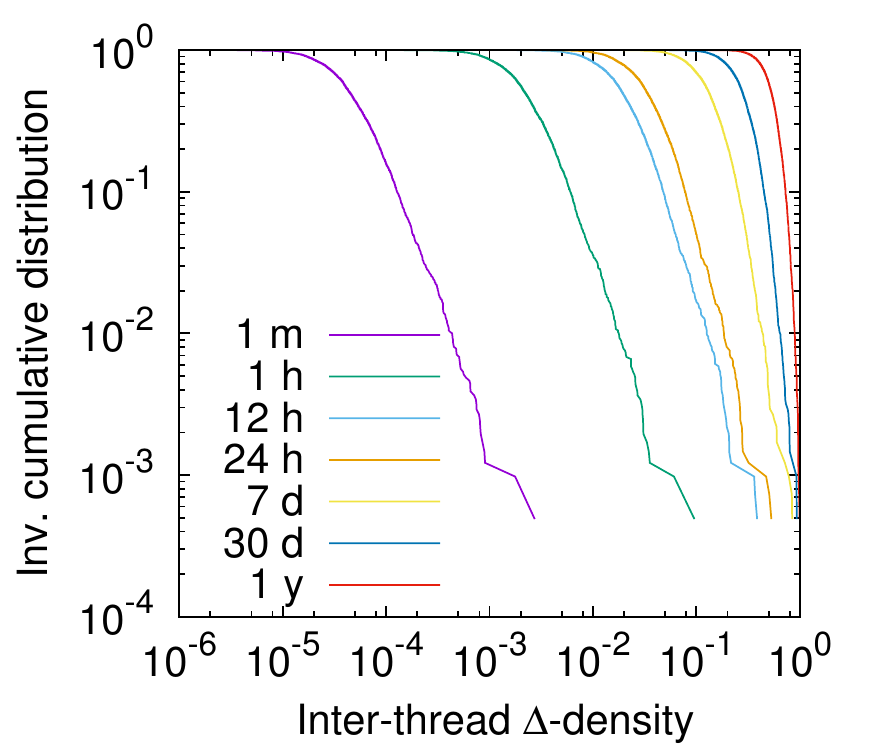}
\caption{Left: Inverse cumulative distributions of values of intra-thread $\Delta$-density for different $\Delta$s. Right: Inverse cumulative distributions of values of inter-thread $\Delta$-density for different $\Delta$s.}
\label{fig:inter_intra}
\end{figure}

In our data, the  inverse cumulative distribution of intra-thread $\Delta$-densities are in Figure~\ref{fig:inter_intra} (left) for several values of $\Delta$ ranging from 1 minutes to 1 year.
For each point on the $x$-axis, the plot gives the proportion of threads in the \ml\ that have an intra-thread $\Delta$-density higher than $x$.
As expected, the higher the $\Delta$ used, the higher the density is.
However, there is no significant change between a $\Delta$ of 7 days and a $\Delta$ of 1 year.

Moreover, these distributions confirm that the interactions within threads are much denser (both structurally and temporally) than in the global mailing-list.
Indeed, the median intra-thread $\Delta$-density ranges from $2.69 \times 10^{-4}$ to $0.28$ while the link stream $\Delta$-density ranges from $1.05  \times 10^{-10}$ to $3.42 \times 10^{-5}$.
The intra-thread $\Delta$-density typically is $10^{5}$ times larger than the global $\Delta$-density.

This shows that threads are indeed dense substreams in our link streams. 

%
%
%
%
%

\section{Relations between threads}

In the previous section, we focused on structural and temporal properties {\em inside} threads, compared to the whole link stream. We now turn to the study of relations {\em between} threads.

\subsection{Inter-thread density}


Let us first study the density of relations between threads in a way similar to above. Given a graph $G=(V,E)$ and a partition $C = \{C_i\}_{i_1..k}$ of $V$ into $k$ communities, the inter-community density is the probability that two nodes chosen at random in two different communities are linked together:


$$
	\delta^{inter}(C_i) = \frac{1}{|C|}\sum_{j, i\ne j}\frac{|\{(u,v)\in E\mbox{ s.t. }u\in C_i\mbox{ and }v\in C_j\}|}{|C_i|\cdot |C_j|}
$$

Again, this notion does not directly make sense in link streams, as threads do not induce a partition of nodes. As a consequence, we introduce the inter-thread $\Delta$-density as the probability that two randomly chosen nodes in different communities are linked together during a time interval of duration $\Delta$ chosen at random during the time duration of both threads.


Let us define the inter-thread substream between a thread $P_i$ and a thread $P_j$: $L_{ij} = (T_{ij}, V_{ij}, E_{ij})$, with $T_{ij} = [$min($\alpha_i,\alpha_j$), max($\omega_i, \omega_j$)$]$, $V_{ij} = V_i\cup V_j$ and $E_{ij} = \{(t,u,v): t\in T_{ij}, u,v\in V_{ij}, (t,u,v)\in E\setminus E_i\cup E_j\}$. In other words, this is the substream containing the links between nodes of $P_i$ or $P_j$ that are not involved in threads $P_i$ and $P_j$. The inter-thread density between $P_i$ and $P_j$ is the $\Delta$-density of $L_{ij}$. In order to obtain the inter-thread $\Delta$-density of $P_i$ to all other threads, we simply average the inter-threads $\Delta$-densities of $P_i$ and all other threads. More precisely:


$$
	\delta^{inter}_{\Delta}(C_i) = \frac{1}{|C|}\sum_{j,i\ne j} \delta_{\Delta}(L_{ij})
$$


In our data, the  inverse cumulative distribution of inter-thread $\Delta$-densities are displayed in Figure~\ref{fig:inter_intra} (right) for different values of $\Delta$.
For each point on the $x$-axis, the plot gives the proportion of threads in the \ml\ that have an intra-thread $\Delta$-density higher than $x$.
Again, larger $\Delta$ correlates with larger $\Delta$-densities.
However, the inter-thread $\Delta$-density does not plateau, even for large values of $\Delta$.
This is natural, since the number of links considered in the computation of the inter-thread $\Delta$-density naturally grows with $\Delta$.

\begin{figure}[!h]
\centering
\subfloat[$\Delta= $1 minute]{
	\includegraphics[width=0.48\linewidth]{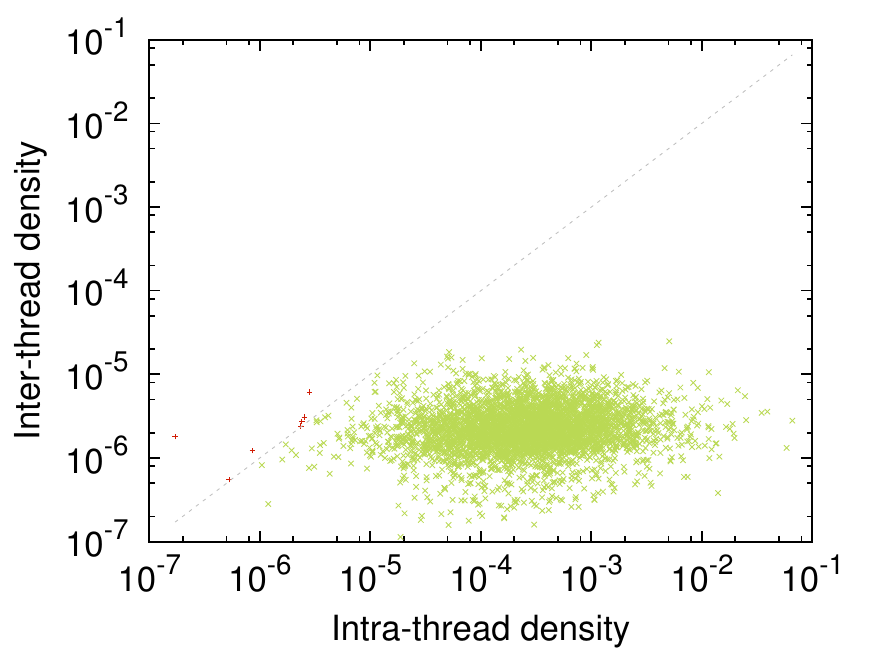}
}\hfill
\subfloat[$\Delta= $1 day]{
	\includegraphics[width=0.48\linewidth]{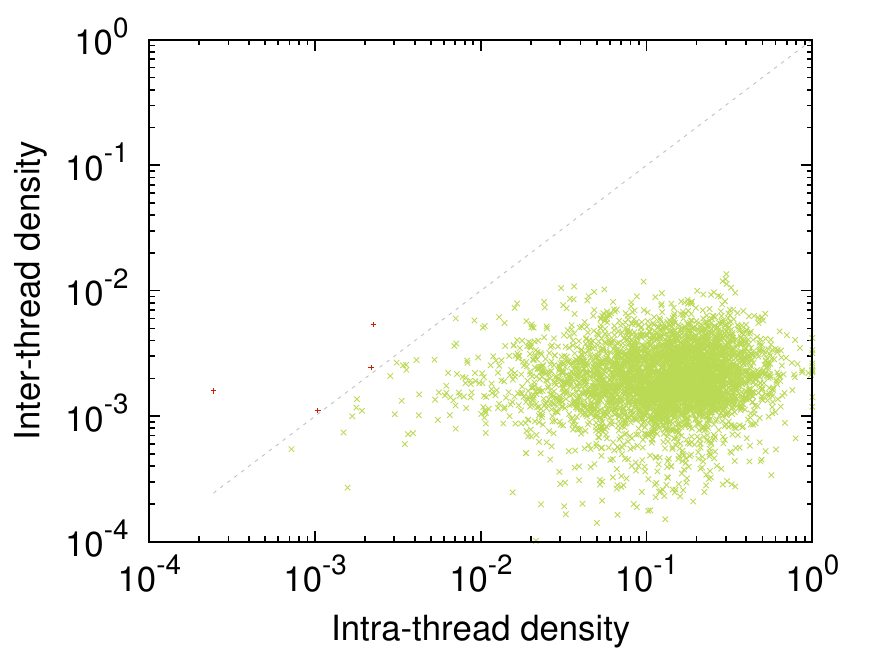}
} \hfill
\subfloat[$\Delta= $1 year]{
	\includegraphics[width=0.48\linewidth]{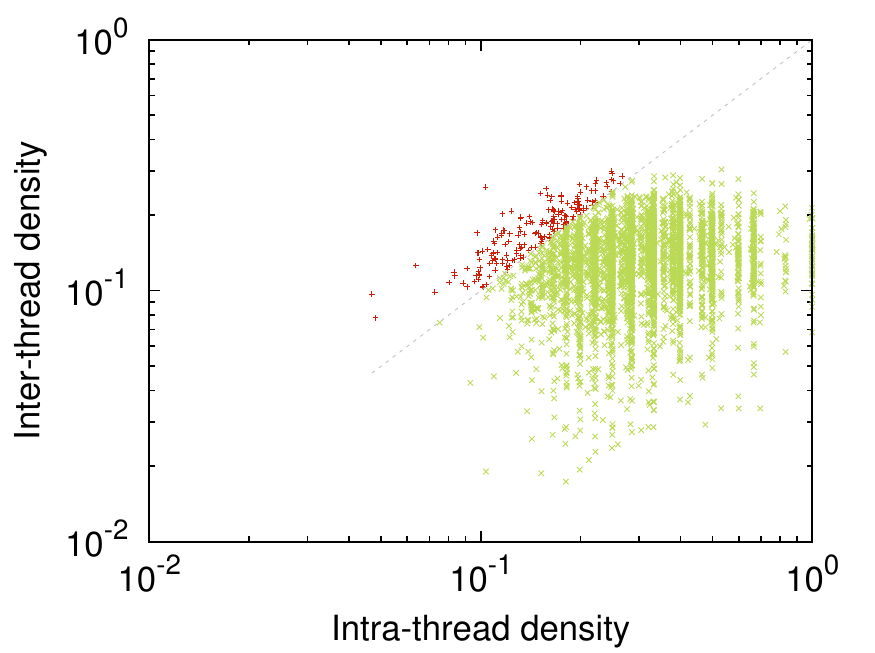}
}
\subfloat[$\Delta= $20 year]{
	\includegraphics[width=0.48\linewidth]{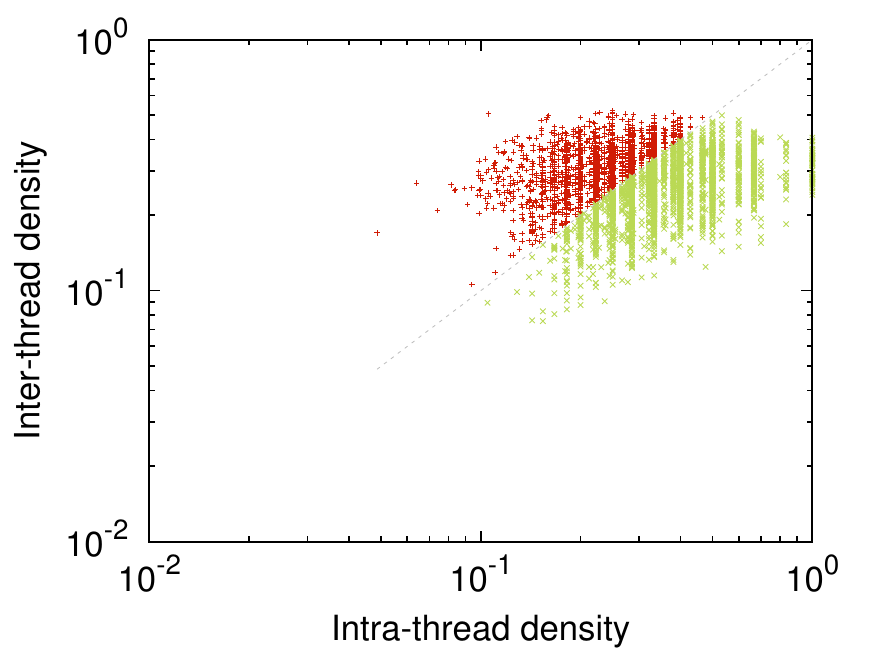}
}
\caption{Correlations between inter- and intra-thread densities for different values of $\Delta$.}
\label{fig:corel_inter}
\end{figure}

In Figure~\ref{fig:corel_inter}, the correlations between the inter- and intra-thread $\Delta$-density are plotted for some values of $\Delta$.
As expected, intra-threads are denser than inter-threads.
This relation holds as $\Delta$ is bigger, even though the difference between inter and intra thread $\Delta$ shrinks. Further experimentation shows that for $\Delta =20$ years, the difference is non-existent. The figure is omitted for brevity. 
This is due to the fact that the bigger the $\Delta$, the less the temporal characteristics of threads are important. 

\subsection{Graphs between threads}
Relations between sub-streams $L_i$, $i=1..k$, may have different forms, and in particular they have a temporal and a structural nature. In order to capture the temporal relations between sub-streams, one may define the temporal overlap graph as follows: $X = (V,E)$ with $V = \{ i, i=1..k \}$ and there is a link $(i,j)$ in $E$ whenever $P_i$ and $P_j$ have a temporal intersection ({\em i.e.} $[\alpha_i,\omega_i] \cap [\alpha_j,\omega_j] \not= \emptyset$). Likewise, one may define the node overlap graph as follows: $Y = (V,E)$ with $V = \{ i, i=1..k \}$ again and there is a link $(i,j)$ in $E$ whenever there is a node $v$ involved in both $P_i$ and $P_j$ ({\em i.e.} there exists a $t$, a $t'$n a $u$ and a $u'$ such that there is a link $(t,u,v)$ in $P_i$ and a link $(t',u',v)$ in $P_j$.

The graphs contain $116999$ nodes (the number of threads) and about $2$ million edges for the temporal overlap graph and $63$ millions for the node overlap graph.
These graphs encode much information about relations between threads.
For instance, the degree of node $i$ in $X$ is the number of threads active at the same time as $P_i$.

\begin{figure}
\centering
	\includegraphics[width=0.49\linewidth]{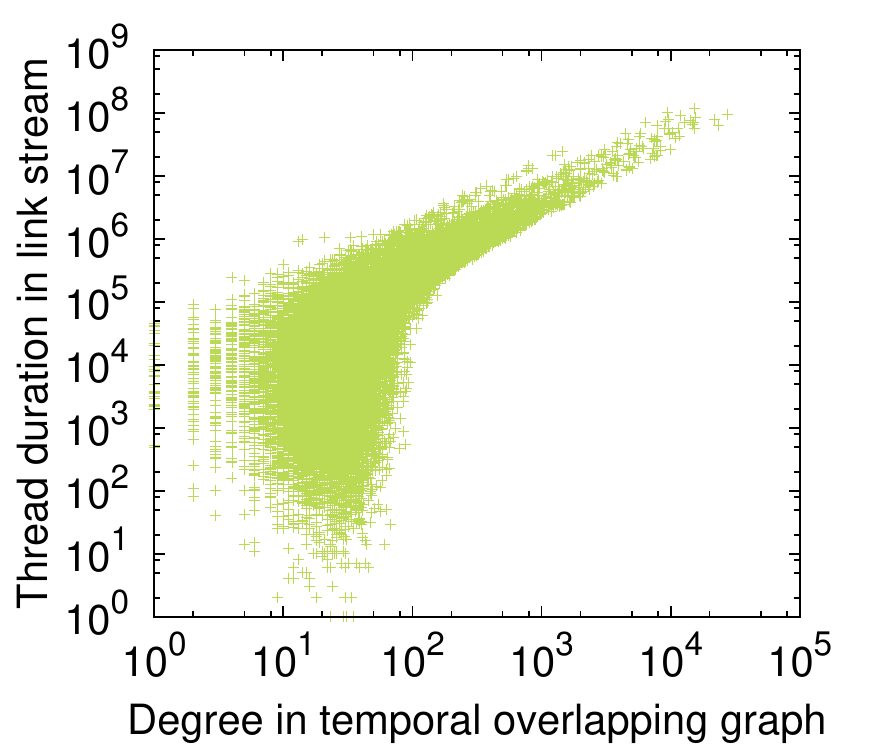}
	\hfill
	\includegraphics[width=0.49\linewidth]{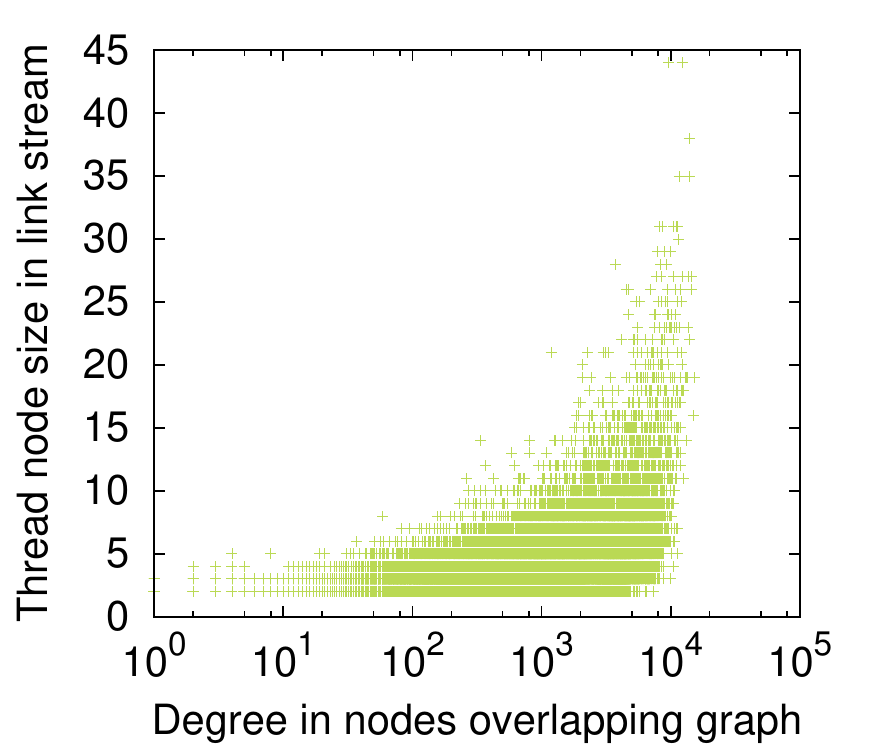}
	\caption{Left: Correlation between the degree in the time overlap graph $X$ and the thread size. Right: Correlation between the degree in the node overlap graph $Y$ and the thread duration.}
	\label{fig:x-y-graphs}
\end{figure}

We display in Figure~\ref{fig:x-y-graphs} (left) the correlations between the degree in $X$ and the thread size. 
There is a clear correlation between the thread duration and the degree in temporal overlap graph when threads have a duration of at least $10^5$s.
Also, it appears that some time up to $10^4$ threads are present simultaneously as reflected by the maximal degree. 

Figure~\ref{fig:x-y-graphs} (right) shows the correlations between the degree in $Y$ and the thread duration.
The correlation is less clear between the thread node size and the degree in the node overlap graph.
However, the trends appears: threads with a lot of participants have a high degree in the graph.
 

\subsection{Quotient stream}

\begin{figure}
\centering
	\includegraphics[width=0.50\linewidth]{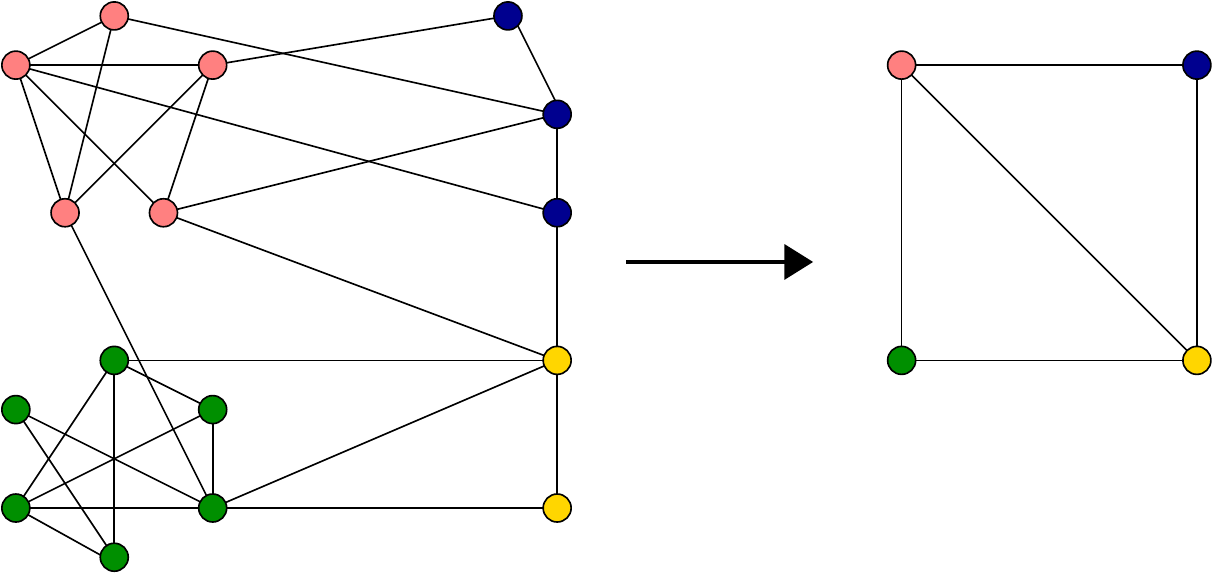}
	\vfill
	\includegraphics[width=0.75\linewidth]{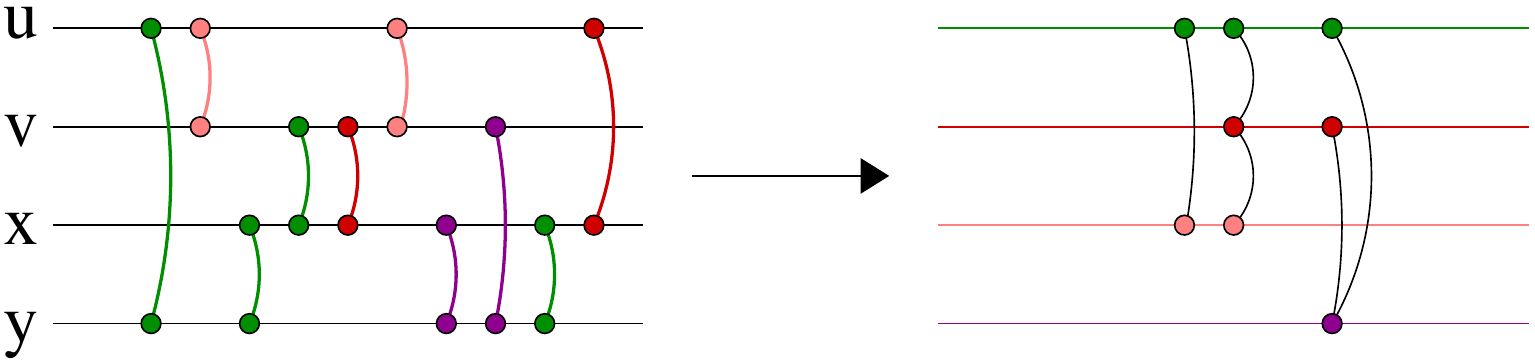}
	\caption{Top: An example of graph exhibiting communities and its corresponding graph quotient. Bottom: An example of link stream with communities and its corresponding quotient stream.}
	\label{fig:graph-quotient}
\end{figure}

The {\em quotient} graph is another key notion for studying the relations between communities in a graph $G = (V,E)$. Given a partition $C = \left\{C_i\right\}_{i=1..k}$ of $V$ into communities, in the quotient graph $\overline{G}$ each node $i$, $i=1..k$, represents community $C_i$ and there is a link between two nodes $i$ and $j$, $i\not= j$, if there is a link between a node in $C_i$ and a node in $C_j$ in $G$. See Figure~\ref{fig:graph-quotient} for an illustration. One may add on each link a weight indicating the number of links between communities. Clearly, the quotient graph captures relations between the communities under concern; for instance, its density indicates up to what point all communities have links between them.

To deepen our understanding of our data, we capture here both temporal and structural nature of relations between sub-streams. We define the quotient stream induced by a partition $P = \left\{P_i = (T_i, V_i, E_i)\right\}_{i=1..k}$ of link stream $L$ as the stream $Q = (T_Q, V_Q, E_Q)$ such that $(P_i,P_j,t)\in E_Q$ if and only if there exists $(u,v,t_1)$ in $E_i$, $(u,v',t_2)$ in $E_i$ and $(u,v'',t)$ in $E_j$ with $t_1 \le t \le t_2$. In other words, there is a node $u$ that has a link within $P_j$ occurring between two of its links in $P_i$. This means that $u$ is involved in the two streams during the same time period.

The quotient stream induced by the threads in our dataset has $12281269$ links and involves $68524$ distinct nodes (i.e. threads). Since our dataset contains $116999$ threads, this implies that $48475$ threads are not in relation with any others.


\begin{figure}
\centering
	\includegraphics[angle=-90, width=0.71\linewidth]{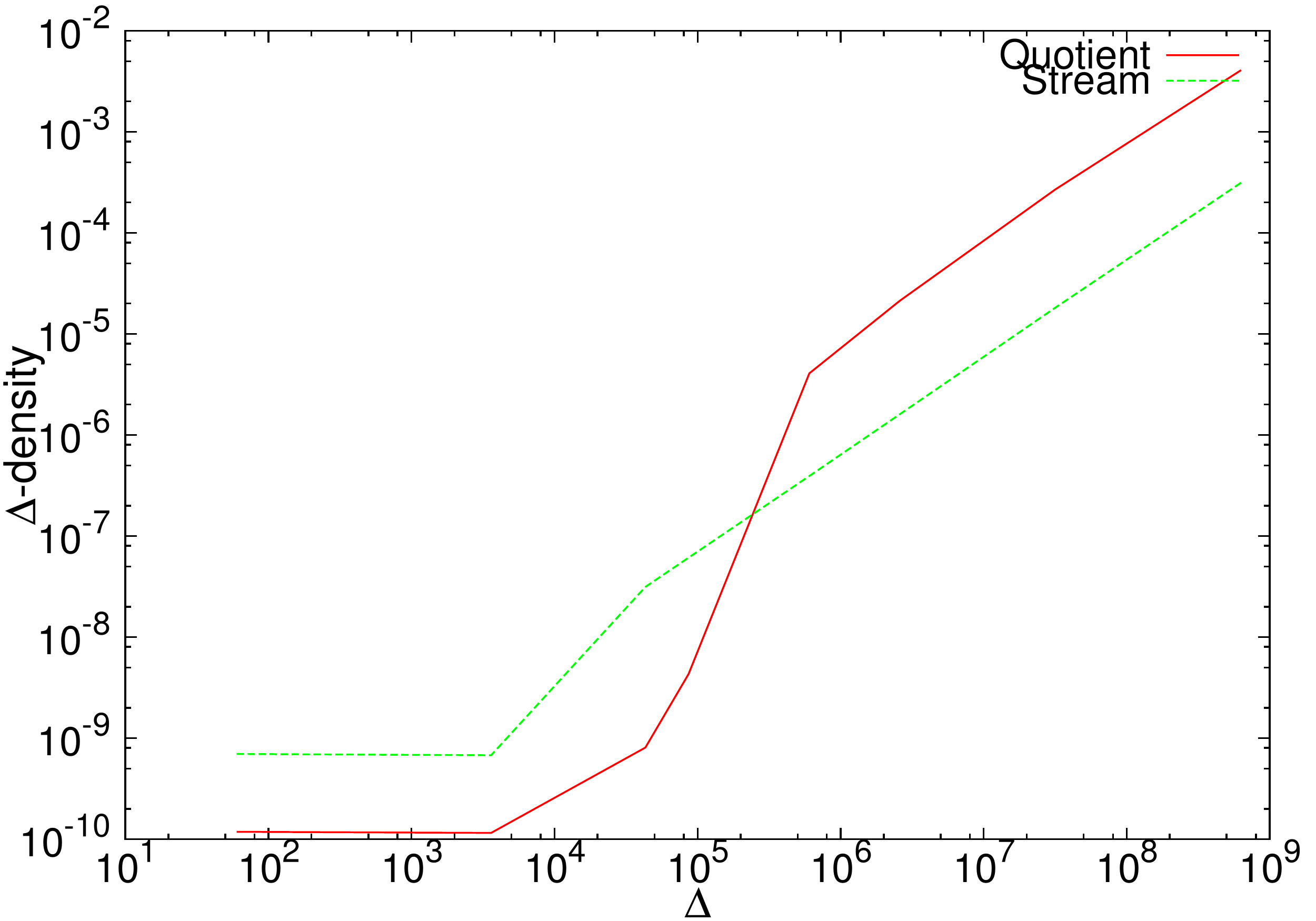}
	\caption{$\Delta$-density of the link stream and the quotient stream as a function of $\Delta$, for $\Delta=1mn, 1h, 12h, 1d, 37d, 30d, 1y$ and $20y$.}
	\label{fig:quotient-stream-density}
\end{figure}

Figure \ref{fig:quotient-stream-density} shows the $\Delta$-density of the quotient stream and the $\Delta$-density of the original stream for different values of $\Delta$. The quotient is not very $\Delta$-dense, i.e. threads are not densely connected together, though it is slightly denser than the stream for large values of $\Delta$. This is comparable to graphs.




%

\section{Conclusion}



Through the prism of link streams, we have studied the email exchanges in the \debianml{} over almost $20$ years. From $\Delta$-density, we define notions of \emph{inter} thread density, \emph{intra} thread density and quotient stream, that are generalizations of the equivalent notions in graphs. We show the relevance of these notions on a real-world dataset of email exchanges.

We have shown that threads in the \ml\ are $\Delta$-dense substreams from a link stream perspective, just like communities are dense subgraphs from a graph perspective.
Moreover, the threads appear to be denser internally than externally which is another feature of communities in graphs.
We also study the relations between threads with the node and temporal overlap graphs.
These graphs reveal the highly temporal and  structural overlapping nature of threads.
However these graphs focus either on time or structure. 
The quotient stream successfully accounts for both aspects, and exhibits similar properties as its graph counterpart.

Though threads are readily identified in the \debianml{} archive, this is usually not the case. Detecting dense substreams loosely interconnected without \emph{a priori} knowledge remains a challenge.

Notice that our approach is dependant of a parameter $\Delta$, that has to be chosen externally. Considering links with durations (i.e. $(b,e,u,v)$, meaning that $u$ and $v$ are interacting continuously from $b$ to $e$) instead of punctual links is a promising direction of work.

\medskip

{\bf Acknowledgments.} This work is supported in part by the French {\em Direction Générale de l’Armement} (DGA), by the Thales company, by the CODDDE ANR-13-CORD-0017-01 grant from the {\em Agence Nationale de la Recherche}, and by grant O18062-44430 of the French program {\em PIA -- Usages, services et contenus innovants}.

\bibliographystyle{plain}
\bibliography{article}

\end{document}